# Excellent thermoelectric performance and impressive optoelectronic properties of Janus monolayer of ZrXY(X=O, S) (Y=S, Se)

Chayan Das, Atanu Betal, Jayanta Bera, Satyajit Sahu

*Department of Physics, Indian Institute of Technology Jodhpur, Jodhpur 342037, India*

## Abstract:

Lower-dimensional TMDC materials are suitable for thermoelectric applications for their specific quantum confinement and being distinct in the density of states (DOS). Here we investigated thermoelectric parameters of the 2D TMDC monolayer of ZrXY ((X=O, S,) (Y=S, Se)) by using Density Functional Theory (DFT) combined with Boltzmann Transport Equation (BTE). We obtained an excellent thermoelectric figure of merit (ZT) of ZrOS, ZrOSe, and ZrSSe at 900 °C. As ZrSSe monolayer showed a ZT value of 0.76 and 0.69 for n-type, and for p-type respectively at 900 °C, so, we investigated new Janus monolayers of ZrOS and ZrOSe, and found a maximum ZT of 0.82 and 0.72 for n-type and p-type, respectively, at 900 °C which is quite impressive. Investigating the electronic property, we found that the ZrOS, ZrOSe, and ZrSSe all possess an indirect bandgap (BG) of 1.88eV, 1.02eV, and 0.74eV respectively. ZrOS and ZrOSe show very high absorption in the ultraviolet (UV) region. However, ZrSSe showed a strong absorption between the red and the yellow region, which suggests that these materials are useful in optoelectronic devices in their wavelength range as well as in thermoelectric applications.

## 1. Introduction:

After the discovery of Graphene 2D materials (TMDC) attracted our attention because of their extraordinary optical [1], electronic [2], thermal [2], properties. Devices based on monolayers of TMDC showed excellent electrical properties with less power consumption; using them even electricity can be generated from waste heat. In this modern era, renewable energy sources are needed as the non-renewable energy sources are depleting very fast. In this work, we investigated low-dimensional TMDC materials using Boltzman transport theory associated with DFT which showed a very high Seebeck coefficient(S). Seebeck coefficient is the measure of potential generated when a temperature difference is introduced between the two ends of the material. For good efficiency or a high ZT value, both the electrical conductivity (σ) and the Seebeck coefficient (S) should be high, along with a lower thermal conductivity ($k = k_{el} + k_{ph}$). The thermal conductivity $k$ has two components, $k_{el}$, and $k_{ph}$, which signifies the contribution coming from electrons and phonons. Generally, TMDC materials show the very good last two properties along with bandgap tunability [3–5]. These layered 2D materials showed excellent tunable optoelectronic properties compared to the bulk [6], which led researchers to use them in electronic [7], optoelectronic [8] applications, as well as thermoelectric [9] and piezoelectric [10] energy generators. TMDCs are the types of materials that can be represented by X-M-X where X is the chalcogenide atom (S,Se,Te) and M is the Transition metal atom (Mo, Sn, W, Zr, Pt, etc.) in hexagonal planes sandwiched between them the chalcogenide by weak Van-der Waal interaction [11]. They can be synthesized easily by exfoliation with scotch tape, Chemical Vapor Deposition (CVD), and thermal deposition method. Mei Zhang et al. synthesized hexagonal $ZrS_2$ monolayer on Boron Niride (BN) and reported maximum mobility of 2300 cm$^2$V$^{-1}$s$^{-1}$ [12]. SD Guo and co-workers investigated the thermoelectric properties of $ZrS_2$ and ZrSSe Janus monolayer and reported a ZT value greater than 0.9 at 600K. The $ZrS_2$ and ZrSSe monolayer were reported with a thermal conductance of 47.8 W/K and 33.6 W/K, respectively. As a result, ZrSSe Janus monolayer showed a higher ZT value compared to $ZrS_2$ monolayer [13]. Similarly, the Janus monolayer of MoSSe also reported with lower thermal conductance compared to $MoS_2$, which results in higher ZT value of MoSSe compared to $MoS_2$[14]. Various experimental and theoretical studies showed that a unique combination of the low value of $k$ and high value of σ of 2D TMDC materials make them suitable materials for thermoelectric applications for converting waste heat into electricity [15–17]. However, still more research on thermoelectric materials is needed to get efficient heat to electricity conversion at near-room temperature. This is the first time we systematically investigated with great detail the electronic and thermoelectric parameters along with optical properties of Janus monolayers of ZrOS and ZrOSe using DFT and BTE. In ZrOS monolayer, the ZT product reaches a maximum of 0.82 for n-type (0.72 for p-type) at a temperature of 900K. Similarly, ZrOSe and ZrSSe Janus monolayer also give a maximum ZT of 0.81 and 0.82 for n-type (0.72 and 0.73 for p-type) respectively. Thus, understanding the physical and chemical properties that improve these materials' thermoelectric efficiency necessitates theoretical investigation of thermoelectric and optical properties. ZrSSe showed the absorption peak in the visible region (635nm); whereas, ZrOS and ZrOSe showed the absorption peak in the UV region (337nm and 376nm). So, these materials can be also be used in photodetector corresponding to their wavelength region.

## 2. Methodology:

We accomplished the first principle calculations using DFT with Vanderbilt ultrasoft pseudopotential [18] and PBE as generalized gradient approximation (GGA)[19] in Quantum Espresso (QE) package. In order to avoid the interaction of two layers, we keep a vacuum of 20 Å between two layers along z-direction, which also helps avoid the periodic boundary condition. The calculation was performed using 15×15×1 k-mesh grid and the geometry was optimized. All the calculations were done with wavefunction energy cutoff of 50Ry and self-consistency was set to $10^{-9}$ Ry. The atoms were relaxed till the force convergent threshold of $3.8 \times 10^4$ Ry was achieved. The phonon dispersion band structure was evaluated with 8×8×1 q-grid using Density Functional Perturbation Theory (DFPT) in QE package. The optical parameters were obtained using the SIESTA package with Time Dependent Density Functional Perturbation Theory (TD-DFPT)[20]. Using Momentum space formulation along with Kramers-Kronig transformation [21], we obtained the imaginary ($\varepsilon_i$) and real ($\varepsilon_r$) parts of dielectric functions. After, that we obtained the absorption coefficient (α), refractive index (η), and extinction coefficient (K) one by one using the following equations.

$$\eta = \left[\frac{\left\{(\varepsilon_r^2 + \varepsilon_i^2)^{1/2} + \varepsilon_r\right\}}{2}\right]^{1/2} \qquad (1)$$

$$K = \left[\frac{\left\{(\varepsilon_r^2 + \varepsilon_i^2)^{1/2} - \varepsilon_r\right\}}{2}\right]^{1/2} \quad (2)$$

$$\alpha = \frac{2K\omega}{C} \quad (3)$$

Here, $\varepsilon_r$, $\varepsilon_i$, $\omega$, and C are real and imaginary parts of dielectric function, frequency and speed of light, respectively. Boltzmann transport equation was used to obtain the Thermoelectric parameters using constant scattering time approximation from BoltzTraP code[22].

$$\sigma_{l,m} = \frac{1}{\Omega} \int \sigma_{l,m}(\varepsilon) \left[-\frac{\Delta f_\mu(T,\varepsilon)}{\Delta \varepsilon}\right] d\varepsilon \quad (4)$$

$$k_{l,m}(T,\mu) = \frac{1}{e^2 T\Omega} \int \sigma_{l,m}(\varepsilon)(\varepsilon-\mu)^2 \left[-\frac{\Delta f_\mu(T,\varepsilon)}{\Delta \varepsilon}\right] d\varepsilon \quad (5)$$

$$S_{l,m}(T,\mu) = \frac{(\sigma^{-1})_{n,l}}{eT\Omega} \int \sigma_{n,m}(\varepsilon)(\varepsilon-\mu) \left[-\frac{\Delta f_\mu(T,\varepsilon)}{\Delta \varepsilon}\right] d\varepsilon \quad (6)$$

Using these equations, we obtained the transport properties. Here, $\sigma_{l,m}$, $k_{l,m}$, $S_{l,m}$, are the electrical conductivity, thermal conductivity, and Seebeck coefficient respectively. Whereas, e, μ, Ω, T are the electron-charge, chemical potential, unit cell volume, and Temperature, respectively. The $k_{ph}$ was obtained using phono3py package combined with QE. A 2×2×1 supercell with 9×9×1 k-mesh was generated, and self-consistent calculations were performed using default displacement of 0.06 Å to obtain the $k_{ph}$. The heat capacity was calculated using Phonopy package combined with QE using 2×2×1 supercell with k-mesh of 9×9×1.

## 3. Result and Discussions:

### 3.1. Properties and stability

ZrXY ((X=O,S,)(Y=S,Se)) monolayer possesses a hexagonal honeycomb structure and belongs to the space group 164 (P-3 m1)[23]. This Janus structure possesses a slightly reduced symmetry because of the loosened symmetry along the central Zr atom. In monolayer, the Zr atom is sandwiched between two layers of X or Y atoms and the unit cell consists of 3 atoms. Each Zr atom is surrounded by 3 X atoms and 3 Y atoms in upper and lower planes. The top view and side view of the monolayer of ZrXY is shown in **Fig1**.

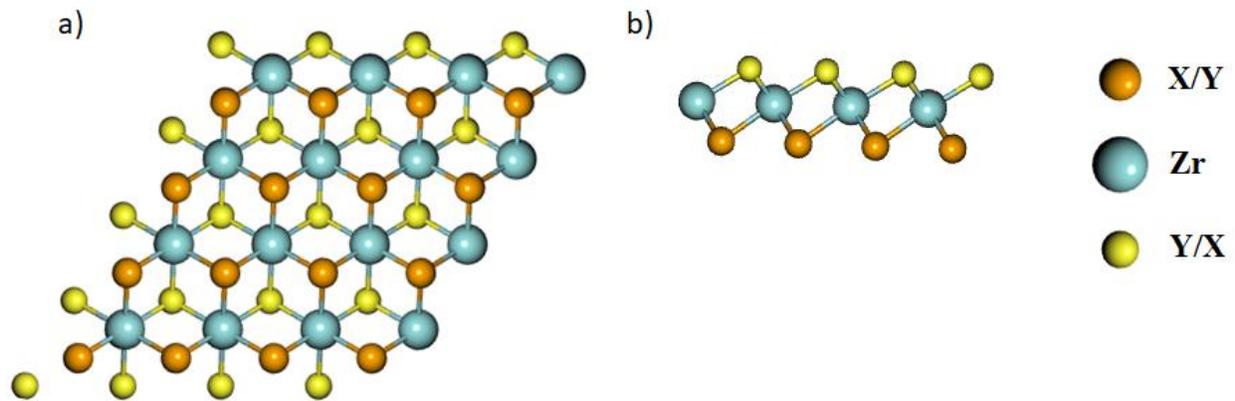

**Fig:1** 4×4×1 supercell of ZrXY monolayer with honeycomb structure: (a) top and (b) side view

By relaxing the unit cells, the obtained lattice constant for ZrOS was found to be a=b=3.45 Å which was in agreement with previously reported results [23], a height of 2.42 Å. For ZrOSe, the parameters were a=b=3.49 Å and 2.55 Å. Similarly for ZrSSe, the obtained lattice constants were a=b=3.75 Å and height of 3.02 Å, which was in agreement with previously reported results [24],. The obtained lattice constants, bond lengths and bond angles for the three different structures were shown in **Table 1**. Here we observed that the bond length of Zr-O in ZrOS, and ZrOSe is almost the same but the Zr-Se bond length in ZrSSe is in between the Zr-S bond length and Zr-Se bond length.

**Table-1:** Estimated lattice constants, bond lengths, heights of monolayer, and bond angles for Janus monolayer of ZrOS, ZrOSe, and ZrSSe

| Structure | a=b(Å) | d(Zr-X) | d(Zr-Y) | Height(Å) | ϴZr-X-Zr | ϴZr-Y-Zr | ϴX-Zr-Y |
|---|---|---|---|---|---|---|---|
| ZrOS | 3.45 | 2.15 | 2.55 | 2.42 | 106.2 | 84.9 | 83.1 |
| ZrOSe | 3.49 | 2.16 | 2.68 | 2.55 | 107.5 | 81.4 | 83.7 |
| ZrSSe | 3.75 | 2.75 | 2.71 | 3.02 | 93.7 | 87.4 | 89.4 |

The cohesive energy gives us information about stability of the material. We also obtained the cohesive energy for these three structures given by the equation: $E_{ch} = \{(E_{zr} + E_x + E_Y) - E_{ZrXY}\}/3$, where $E_{ZrXY}$, $E_{Zr}$, $E_x$, and $E_Y$ are the energy of Janus monolayer of ZrXY, the energy of single Zr atom, the energy of single X atom and the energy of single Y atom. The cohesive energy per atom obtained for ZrOS, ZrOSe, and ZrSSe monolayer was 7.93eV, 7.62eV, and 5.62eV which agrees with the previously reported results of the 2D monolayer of $SiS_2$ and $SiSe_2$[25].

To check the structural stability, we calculated the phonon dispersion curve of ZrOS, ZrOSe, and ZrSSe, respectively, along the high symmetry points K-Γ-M-K as shown in **Fig 2**, and observed no imaginary frequency, which confirms the thermodynamical stability of all these structures. Since the unit cell of the monolayer contains three atoms, there are total nine vibrational modes; among them first three are called the acoustic, and the other six are called the optical modes. The lower three branches correspond to the acoustic vibrational modes, and they are in-plane longitudinal acoustic (LA) mode, transverse acoustic (TA) mode, and out of plane mode (ZA). The upper six modes are optical modes. The LA and TA modes show some linear behavior near k=0, but ZA mode shows quadratic behavior.

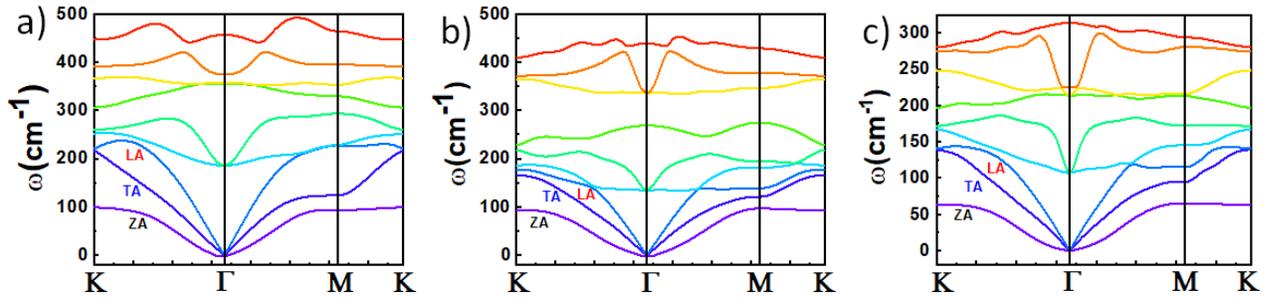

**Figure 2:** The phonon dispersion plot of Janus monolayer of a) ZrOS b) ZrOSe c) ZrSSe

### 3.2. Electronic Properties:

The band structure of Janus monolayer of ZrOS, ZrOSe, and ZrSSe was calculated along K-Γ-M-K path within an energy range from -4 eV to 4 eV. **Fig 3 shows the** electronic band structure of ZrOS, ZrOSe, and ZrSSe with corresponding BGs of 1.88eV, 1.02eV, and 0.66 eV. The ZrSSe BG we obtained matched well with previous work [24]. All three band structures show an indirect bandgap (BG). The valance band maxima (VBM) is situated at Γ point, but the conduction band minima (CBM) for ZrOS and ZrOSe is situated between Γ and M, and for ZrSSe the CBM is found at M point.

The DOS along with some Local density of states (LDOS) for different orbitals, for the monolayer of ZrOS, ZrOSe and, ZrSSe is shown in **Fig 4a, b, and c**. For ZrOS and ZrOSe monolayer the contribution for VBM is mainly from the $p_z$, $p_x$ and $p_y$ orbitals of S and Se atoms and CBM is mainly contributed by $d_{z2}$, $d_{zx}$ and $d_{zy}$ of Zr atom as shown in the Projected density of states (PDOS) in **Fig 4d and e**. For ZrSSe the main contribution towards VBM is from $p_x$ and $p_y$ of S atom and $p_y$ of Se atom, $p_x$ contribution is less but $p_z$ contribution is negligible for Se atom and the contribution towards CBM is mostly from $d_{z2}$, $d_{x2-y2}$ and $d_{xy}$ of Zr atom as shown in **Fig 4f**. Unlike ZrOS and ZrOSe the $p_x$, $p_y$ in ZrSSe are not degenerate. The BG of ZrOS is the highest and of ZrSSe is the lowest, and ZrOSe is in between them. It is observed that for all the monolayer, the O, S, and Se contribute more towards the valance band and Zr contributes more towards the conduction band.

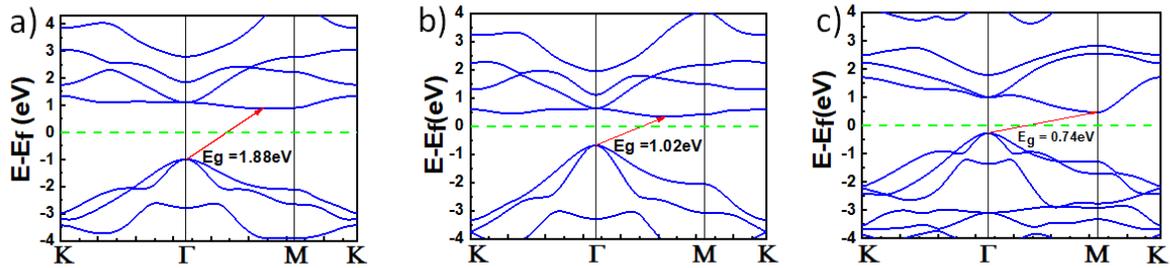

**Figure 3:** The Band structure of Janus monolayer of a) ZrOS, b) ZrOSe, and c) ZrSSe. The red arrow signifies the BG, and the Fermi level is shown by green dotted lines.

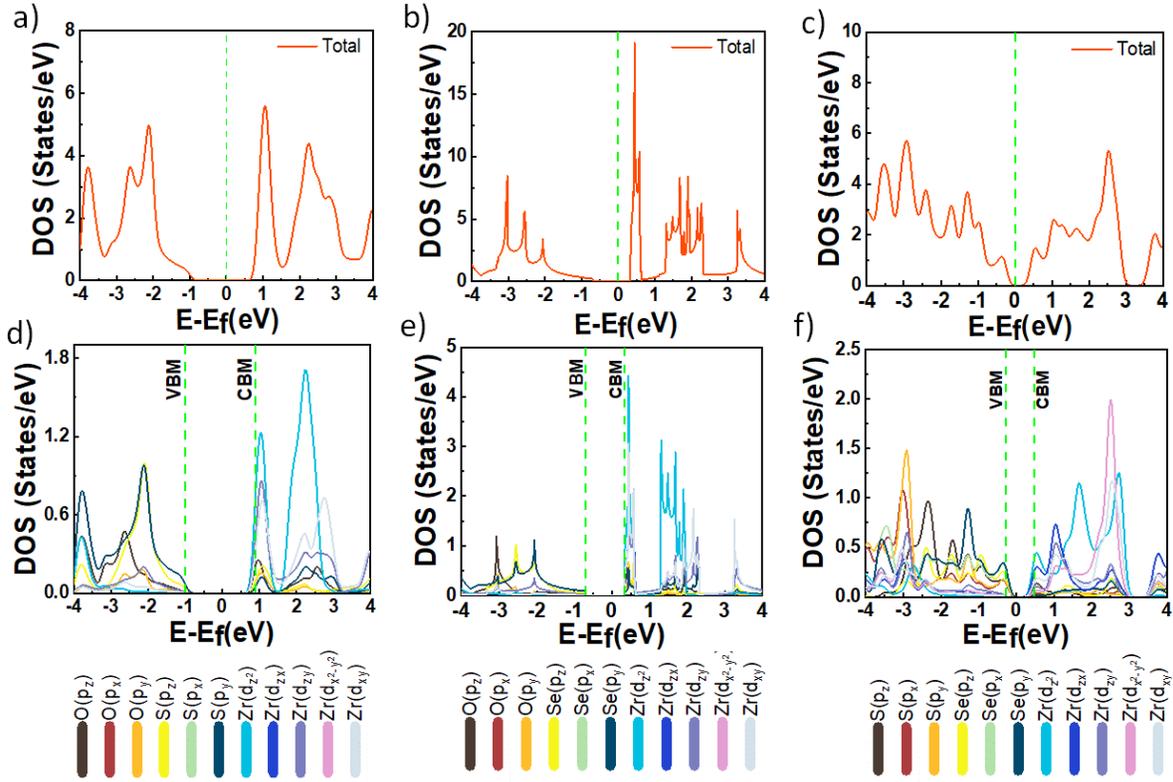

**Fig 4:** DOS of monolayer of a) ZrOS, b) ZrOSe and c) ZrSSe. PDOS of monolayer of d) ZrOS, e) ZrOSe and f) ZrSSe. Here Different color shows different Orbitals.

### 3.3. Optical properties:

Optoelectronic device applications require the calculation of optical properties. The optical properties were investigated along the perpendicular direction of the plane. The $\varepsilon_r$ is obtained using Kramers-Kronig Transformation, and the $\varepsilon_i$ of dielectric function was calculated using momentum space formulation using proper matrix elements. The dielectric functions were plotted as a function of energy of photon in **Fig 5a and b.** The $\varepsilon_i$ shows peak at 2.54 eV for ZrOS, at 2.39eV for ZrOSe and at 0.22eV for ZrSSe. This can be explained from the band structure, ZrSSe has the lowest BG, and ZrOS (1.14 eV greater than ZrSSe) has the largest BG, and in between lies the BG of ZrOSe (0.28eV greater than ZrSSe), so ZrOSe gets larger blue shift compared to ZrSSe, and the blue shift for ZrOS is largest among them. The secondary peaks were found at 5.17eV, 5.24 eV, and at 6.44 eV for ZrOS, ZrOSe, and ZrSSe, respectively. Negative peaks were found at 6.00eV, 3.45eV and 1.35eV for ZrOS, ZrOSe, and ZrSSe, respectively for the $\varepsilon_r$. The absorption spectra were shown in **Fig 5c,** and the peaks were found at 3.75eV, 3.30eV, and 1.80eV with absorption coefficient (α) 7.33×10$^5$/cm, 7.73×10$^5$/cm, and 4.83×10$^5$/cm for ZrOS, ZrOSe, and ZrSSe respectively which are in order of SnI$_2$ and SiI$_2$ [26]. Some secondary peaks are also observed near 6 eV for ZrOS and ZrOSe and near 8eV for ZrSSe with an even higher absorption coefficient, which indicates the use of these materials as UV photodetector. Only ZrSSe can be used in photovoltaic applications in the visible region. The variation of refractive index (η) with energy is shown in figure **Fig 5d.** The refractive index is at zero energy is 3.82, 4.35, and 11.41 for ZrOS, ZrOSe, and ZrSSe, respectively. The secondary peaks are found at energy 1.19eV and 4.8eV for ZrOS and at relatively high energy of 4.94eV and 6.89eV for ZrOSe and 5.24eV and 10.12eV for ZrSSe. After 15eV, the refractive index of ZrOS and ZrOSe is almost the same, but ZrSSe obtained some lower values. As observed from many previous works [25–27], the relaxation time for 2D materials and ZrSSe [28] lies in the order of 10$^{-14}$s.

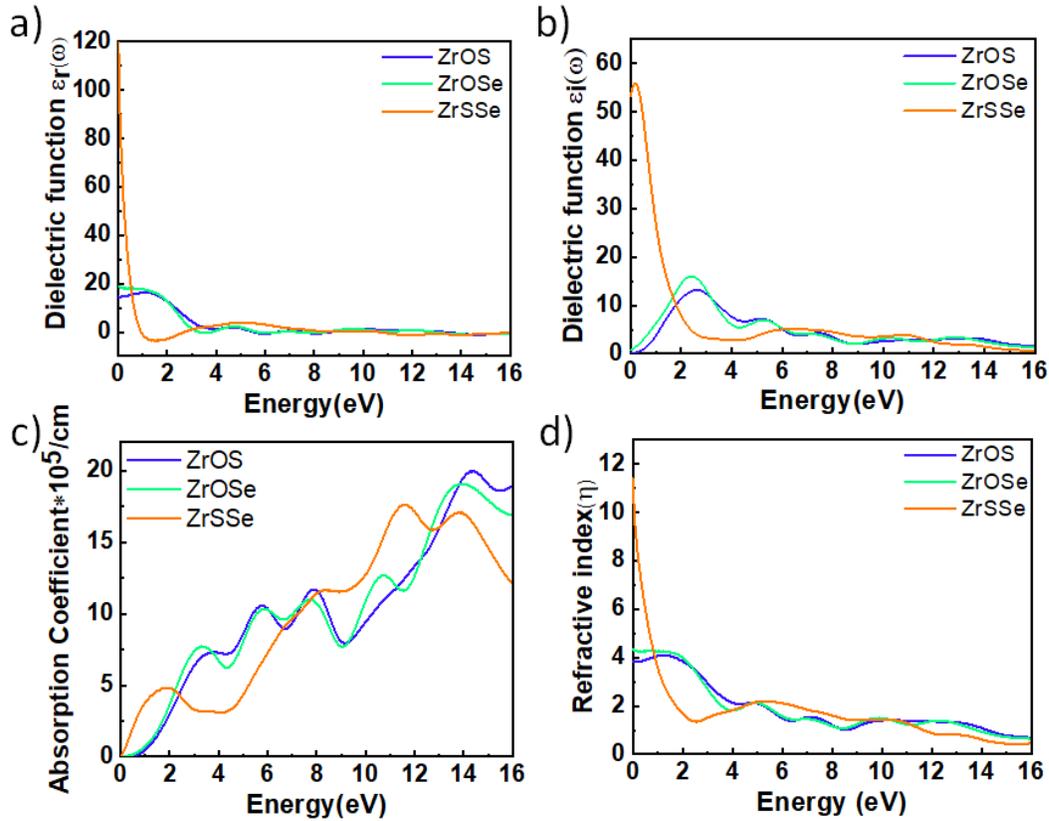

**Fig 5:** Optical properties a)$\varepsilon_r$ , b) $\varepsilon_i$ , c) α, and d) η as a function of energy are shown.

### 3.4. Thermoelectric Properties:

The plot of S with respect to µ at 300K, 600K, and 900K for ZrOS Janus monolayer is shown in **Fig 6a.** The maximum S obtained is 2993 µV/K for n-type carriers (µ>0) and 2764 µV/K for p-type carriers (µ<0) for ZrOS monolayer at 300K. A decrease of S with an increase in temperature (T) was observed. The plot of relaxation time-scaled electrical conductivity ($\sigma/\tau$) with respect to µ is shown in **Fig 6b.** There is not much change in ($\sigma/\tau$) with T. The n-type ZrOS possesses much lower electrical conductivity. The plot of the relaxation time-scaled power factor (PF=$S^2\sigma/\tau$) with respect to µ is shown in **Fig 6c.** The highest PF was obtained for ZrOS monolayer for n-type carriers (33.61×10$^{10}$ W/m$^2$Ks) at 600K and for p-type carriers (23.24×10$^{10}$ W/m$^2$Ks) at 900K. So it is clear that doping with n-type gives more efficiency than that with p-type in thermoelectric applications. For ZrOSe and ZrSSe monolayer the variation of $\sigma/\tau$ and $S^2\sigma/\tau$ with respect to µ is shown in **Fig 6e,f and 6h,i**. The highest power factor we obtained for ZrOSe monolayer is 27.22 ×10$^{10}$ W/m$^2$Ks for n-type carriers at 600K, for ZrSSe monolayer the heighest obtained value is 57.55 ×10$^{10}$ W/m$^2$Ks for n-type carriers at 900K which is the best among all the structures. The highest values of S, ($\sigma/\tau$), ($S^2\sigma/\tau$) and for all the three monolayers are listed in **table 2**.

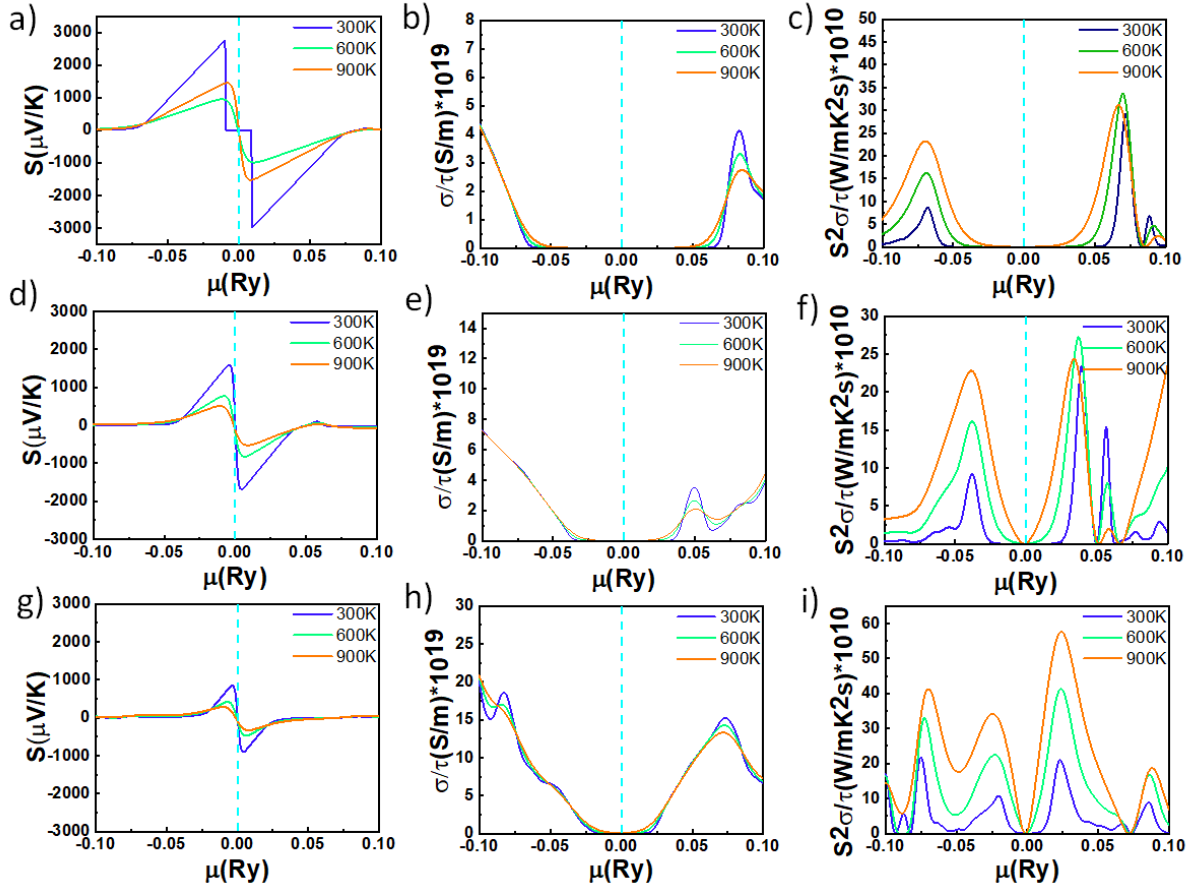

**Fig 6:** Plot of thermoelectric parameters with respect to µ for ZrOS monolayer a), b), and c), for ZrOSe monolayer d), e), and f), and for ZrSSe monolayer g), h), and i).

**Table 2:** Calculated maximum a) S, b) σ/τ, and c) S²σ/τ for Janus monolayer of ZrOS, ZrOSe, and ZrSSe.

| Crystal | Maximum Seebeck Coefficient(S) (µV/K) | | Maximum Conductivity ×10$^{19}$(σ/τ) (S/ms) | | Maximum Power Factor ×10$^{10}$ (S$^2$σ/τ) W/m$^2$Ks | | Thermoelectric figure of merit (ZT) | |
|---|---|---|---|---|---|---|---|---|
| | p | n | p | n | p | n | p | n |
| ZrOS | 2764.14 | 2993.63 | 6.17 | 4.11 | 23.24 | 33.61 | 0.72 | 0.82 |
| ZrOSe | 1596.43 | 1685.44 | 13.51 | 3.51 | 22.75 | 27.22 | 0.72 | 0.81 |
| ZrSSe | 838.46 | 903.56 | 18.56 | 15.23 | 41.27 | 57.56 | 0.69 | 0.76 |

As S is directly proportional to BG, we found the highest S for ZrOS monolayer and lowest S for ZrSSe similar to BG values, i.e., ZrOS has the heighest and ZrSSe has the lowest BG value. And the ZrOSe monolayer, showed a value of S in between the ZrOS and ZrSSe monolayer, which is analogous because ZrOSe has the intermediate BG value with respect to ZrOS and ZrSSe monolayer as discussed earlier. Also, it is clearly seen that for all three structures the n-type doped materials show better thermoelectric properties than that of p-type doped material.

### 3.5. Lattice Thermal Conductivity ($\kappa_{ph}$):

The variation of $\kappa_{ph}$ due to phonon with respect to T for monolayer of ZrOS, ZrOSe and ZrSSe is shown in **Fig 7a, 7d and 7g.** Janus monolayer of ZrSSe showed the lowest $\kappa_{ph}$ (0.666 W/mK) at 300 K, ZrOS and ZrOSe also showed significantly low $\kappa_{ph}$ of 0.963 W/mK and 0.921 W/mK at 300K respectively, which are lower than $Bi_2Te_3$ (1.6 W/m.K) [29], $MoS_2$ (34.5 W/m.K) [30], $WS_2$ (72 W/m.K) [17], PbTe (2.2 W/m.K) [31], $SnS_2$ (15.85 W/(m.K)) [2], $SiSe_2$ (15.85 W/(m.K)) [25] $ZrS_2$ [28], $HfS_2$ (2.84 W/m.K) and $HfSe_2$ [32–34]. A decrease in $\kappa_{ph}$ for all three structures with an increase in T is observed. The ZrSSe monolayer possesses lower thermal conductivity compared to ZrOS and ZrOSe monolayer. The change of phonon lifetime (τ) and group velocity (Gv) as a function of frequency for ZrOS and ZrOSe and ZrSSe monolayers is shown in **Fig7b, 7e, and 7h** respectively. The maximum phonon life time is found to be nearly same for all the three Janus monolayers which is about 1.0 ps which is quite low compared to $MoS_2$ and $WS_2$ [35,36]. The maximum phonon group velocity obtained for ZrOS monolayer is about 7.1 Km/s (**Fig 7c**) which is due to out of plane acoustic mode. Similarly, the maximum obtained Gv for ZrOSe is about 6.1 Km/s (**Fig 7f**) that is due to out of plane acoustic mode. However, a comparatively low Gv about 4.4 Km/s (**Fig 7i**) is obtained for ZrSSe monolayer due to longitudinal acoustic mode. As the $\kappa_{ph}$ is proportional to the Gv, and τ, the change in $\kappa_{ph}$ of the three Janus monolayers can be explained easily. As the τ for the three Janus monolayers are nearly same, the Gv plays a crucial role for determining the $\kappa_{ph}$. The highest Gv is obtained for ZrOS monolayer, the lowest Gv is obtained for ZrSSe monolayer, and the Gv of ZrOSe monolayer lies in between the ZrOS and ZrSSe monolayers. The $\kappa_{ph}$ also shows a similar phenomenon i.e., highest for ZrOS, lowest for ZrSSe, and some intermediate value for ZrOSe monolayer.

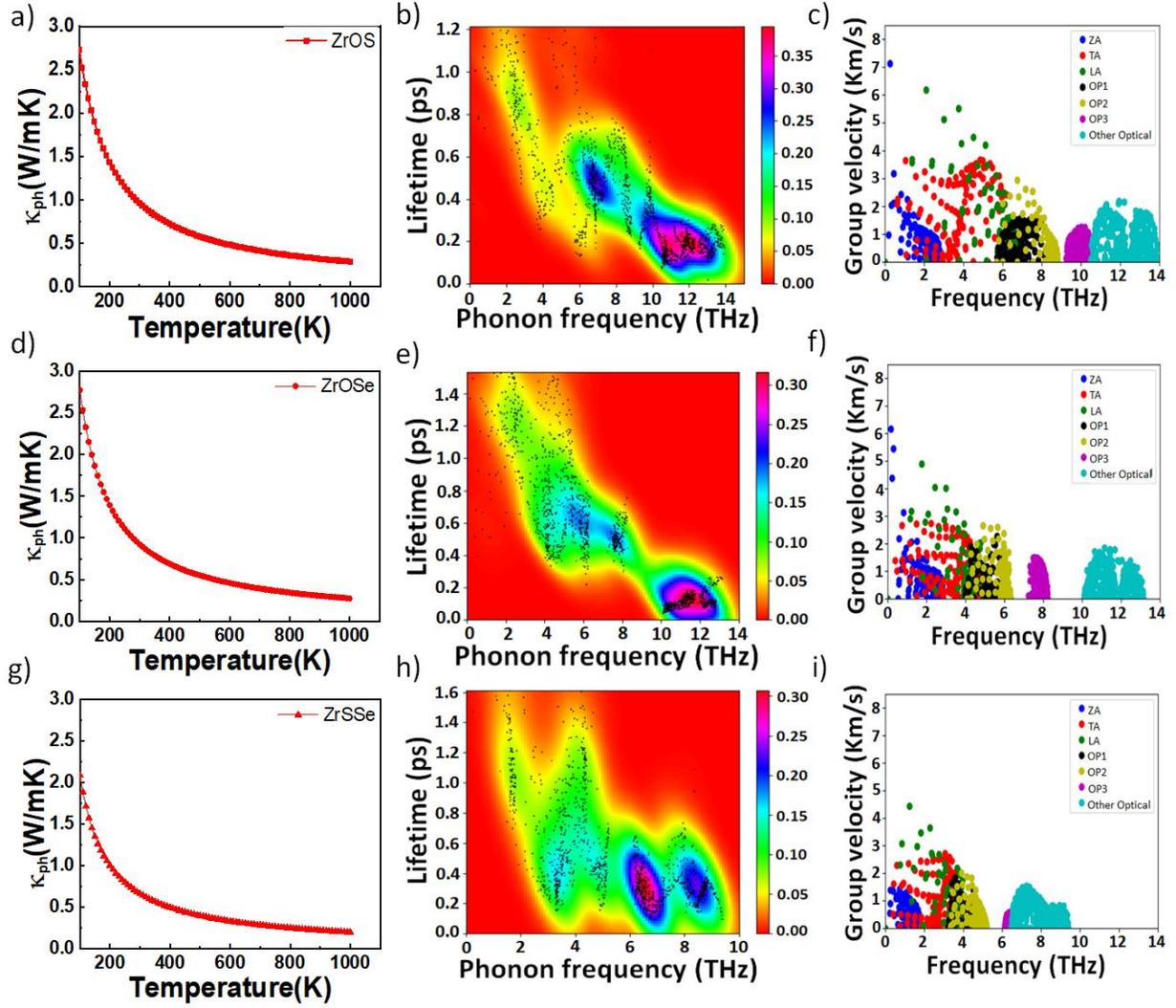

**Fig: 7** Plot of $\kappa_{ph}$, phonon lifetime and group velocity (Gv) with respect to Temperature(K) of monolayer ZrOS, ZrOSe, and ZrSSe.

### 3.6. Thermoelectric figure of merit (ZT)

This parameter ZT signifies the efficiency along with the quality of a material. A high ZT value requires a high value of σ along with a much low value of $k$ ($k_{el} + k_{ph}$). The Thermoelectric figure of merit (ZT) is defined as

$$ZT = \frac{S^2 \sigma T}{\kappa_{el} + \kappa_{ph}} \qquad (7)$$

Where S, T, σ, $k$ and $k_{ph}$ are same as defined earlier. The calculated ZT product is plotted as with respect to μ for Janus monolayer of ZrOS, ZrOSe, and ZrSSe at 300K, 600K, and 900K which is shown in **Fig 8.** Among the three structures, ZrOS and ZrOSe monolayer show the highest ZT value i.e., 0.82 for n-type for both and 0.72 and 0.71 for p-type, respectively. The ZrSSe monolayer showed comparatively low ZT value, i.e., 0.76 for n-type and 0.69 for p-type. All the ZT values for different structures indicate better performance compared to previously reported popular TMDC structures like $MoS_2$, $WS_2$, $HfS_2$, and $HfSe_2$ monolayers. For all the three structures, the n-type monolayer showed a higher ZT value which again represents the effectiveness of doping with n-type compared to p-type. The S of ZrOS is significantly high rather than ZrOSe, but the σ of ZrOSe is significantly higher, so the power factor remains almost same and as their $k_{ph}$ is almost same, the ZT product becomes almost the same. For ZrSSe the S is the lowest, but σ is the highest, and $k_{ph}$ is also the lowest, so the ZT product becomes reasonable.

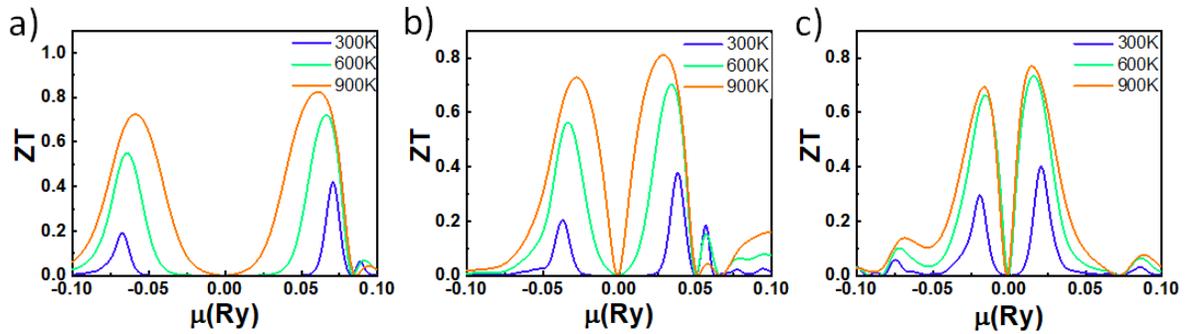

**Fig 8:** The variation of ZT with respect to μ at different temperatures(T) for Janus monolayer of a) ZrOS, b) ZrOSe and, c) ZrSSe.

## 4. Conclusions:

We replaced Sulphur(S) and Selenium (Se) by Oxygen (O) in monolayer of $ZrS_2$ and $ZrSe_2$ and one Sulphur(S) with Selenium (Se) atom in $ZrS_2$ monolayer to create the Janus monolayer structures. We calculated the electronic, optical, and thermoelectric properties using DFT and BTE. The dynamical stability is verified by phonon dispersion curves with no imaginary frequency, which implies that these structures are stable. Though the ZT and $k_{ph}$ for ZrOS and ZrOSe are almost same, but for ZrSSe the ZT along with $k_{ph}$ is less. Though the $k_{ph}$ of ZrOS and ZrOSe is comparatively high, but the S and σ compensate for the power factor and ZT product.

The maximum ZT product is 0.82 for n-type and 0.72 for p-type for Janus monolayer of ZrOS monolayer at 900K. We found that Janus monolayer of ZrOS and ZrOSe show a higher ZT product than ZrSSe, so we can conclude that doping with Oxygen can significantly increase the thermoelectric performance in TMDC. So, ZrOS and ZrOSe can be used for next generation thermoelectric devices to convert waste heat into usable electricity.

**Acknowledgements**